 \documentclass[final,5p,times,twocolumn,authoryear]{elsarticle}

\usepackage{amssymb}
\usepackage{lipsum}
\usepackage{siunitx}
\usepackage{amsmath}
\usepackage{hyperref}
\usepackage{fontenc}
\usepackage[utf8]{inputenc}

\journal{High Energy Astrophysics}

\begin{document}

\begin{frontmatter}



\title{Faint supernovae and hyper-runaway white-dwarfs from single He-detonation in double HeCO-white-dwarf mergers}


\author[1,2]{Hila Glanz}
\ead{glanz@campus.technion.ac.il}
\affiliation[1]{Technion - Israel Institute of Technology,
Haifa, 3200002, Israel}

\affiliation[2]{TAPIR, Walter Burke Institute for Theoretical Physics,
California Institute of Technology, Pasadena, California 91125, USA}

\author[1]{Hagai B. Perets}

\author[3]{Aakash Bhat}
\affiliation[3]{Institut für Physik und Astronomie,
Universität Potsdam, Karl-Liebknecht-Str. 24--25,
14476 Potsdam-Golm, Germany}

\begin{abstract}
We present three-dimensional hydrodynamical simulations of mergers between low-mass hybrid HeCO white dwarfs (WDs), offering new insights into the diversity of thermonuclear transients. Unlike previously studied mergers involving higher-mass HeCO WDs and CO WDs, where helium detonation often triggers core ignition, our simulations reveal incomplete helium shell detonations in comparable-mass, lower-mass WD pairs.  The result is a faint, rapidly evolving transient driven by the ejection of intermediate-mass elements and radioactive isotopes such as \(^{48}\text{Cr}\) and \(^{52}\text{Fe}\), without significant \(^{56}\text{Ni}\) production. These transients may be detectable in upcoming wide-field surveys and could account for a subset of faint thermonuclear supernovae. Long-term evolution of the merger remnant shows that high-velocity PG-1159-type stars might be formed through this scenario, similar to normal CO-CO white dwarf mergers. This work expands our understanding of white dwarf mergers and their implications for nucleosynthesis and stellar evolution.
\end{abstract}




\end{frontmatter}




\section{Introduction} \label{sec:intro}
The study of thermonuclear supernovae originating from white dwarf (WD) mergers has grown significantly in recent years, revealing the diversity of explosive outcomes from double-degenerate systems. Previous studies have explored a range of outcomes in double-degenerate mergers, particularly those involving carbon-oxygen (CO) and helium-rich white dwarfs. 

The double-detonation scenario has long been considered a promising pathway for thermonuclear explosions of sub-Chandrasekhar mass white dwarfs. Originally proposed in the 1980s \citep{Nomoto1980,WoosleyWeaverTaam1980,Livne1990}, the model envisions a helium shell detonation on the surface of a carbon–oxygen (CO) white dwarf that drives converging shocks into the core, igniting carbon and producing a Type Ia-like explosion. While early implementations with massive helium shells predicted strong signatures of iron-group elements at high velocities—largely inconsistent with normal Type Ia spectra—later studies showed that much thinner helium shells can trigger core ignition of massive CO-WDs while producing ejecta consistent with observations \citep{Fink2010,Sim2010,ShenBildsten2009,Woosley2011}. These developments have raised interest in double detonations as a viable progenitor channel for both normal SNe Ia and a variety of peculiar thermonuclear transients.

In recent years, attention has increasingly turned to hybrid helium–carbon–oxygen (HeCO) white dwarfs as natural sites for double detonations. These WDs, predicted by stellar evolution models and identified in binary population synthesis studies, possess extended helium mantles surrounding CO-rich cores \citep{Zen+18}. Their structure makes them particularly prone to surface helium ignition during accretion or mergers. Depending on the total mass and the efficiency of shock convergence, CO-WDs merging with HeCO WDs can potentially produce normal Type Ia explosions when (CO-WD accreting Helium) core ignition is achieved, or faint, rapidly evolving transients dominated by helium-burning products when only the shell detonates \citep{Bildsten_2007,2010PeretsNature,Waldman2011,Per+19,Gro+21,2021PakmorZenatiHybridIgnition,Zen+23,Glanz2025HVWD}. This duality highlights the central role of hybrid HeCO WDs in bridging the classical double-detonation framework with the growing diversity of observed fast and faint thermonuclear supernovae, directly motivating the present study.

While mergers of massive ($>0.8$ M$_\odot$) CO WDs with hybrid HeCO WDs have been studied more extensively, showing double detonation and the production of luminous, possibly normal type Ia SNe, 
helium shell detonations without core ignition which may occur on lower-mass WDs have been proposed as progenitors of fast, faint thermonuclear transients, Ca-rich SNe \citep{2010PeretsNature, Waldman2011}. Other simulations have also investigated the conditions required for shell ignition in mergers or AM CVn systems \citep{Brooks2015, Bauer2017}. However, the potential for thermonuclear explosions from lower-mass HeCO white dwarf systems remains less explored, particularly in cases where helium shell detonations produce elements other than the familiar \(^{56}\text{Ni}\), resulting in fainter, faster transient events.

Using 2D simulations, \cite{Zen+23} have shown that low-mass HeCO hybrid WDs mergers with low mass CO WDs can explain the origin of Ca-rich SNe, and successfully reproduce their observed properties.
\cite{2021PakmorZenatiHybridIgnition} investigated the merger of a high-mass HeCO WD with a CO WD, where helium accretion triggers a robust shell detonation, often igniting the underlying carbon core. This sequence leads to high-energy supernovae with prominent radioactive elements. In our previous work \citep{Glanz2025HVWD}, we demonstrated that mergers between two high-mass HeCO WDs can result in a double detonation, producing a faint supernova and ejecting the disrupted secondary as a hypervelocity WD.  

In this study, we extend our previous three-dimensional (3D) analysis to the case of a merger of two lower-mass HeCO WDs, focusing on cases where helium shell detonations do not trigger secondary core ignition. These mergers produce unique isotopic abundances, including iron-group elements such as  \(^{48}\text{Cr}\) and \(^{52}\text{Fe}\), but little to no \(^{56}\text{Ni}\). These result in subluminous, short-lived thermonuclear transients that are underrepresented in existing models, though have been seen in one or two-dimensional (1D/2D) models of low-mass HeCO WDs explosions \cite{Bildsten_2007,Waldman2011,Zen+23}

Using the same 3D hydrodynamical tools and methods as we used in \cite{Glanz2025HVWD}, including the moving mesh code \texttt{ArepoREPO}, coupled with an extensive nuclear reaction network, we model this lower-mass merger scenario in detail. Our simulation reveals that helium shell detonations in these systems lead to the ejection of intermediate-mass elements, while the primary WD remains intact. This presents a rare observational signature, potentially detectable by upcoming high-cadence surveys, which may open a new window on the nature of low-luminosity transients in the universe.

\section{Methods} \label{sec:methods}

To study the merger outcomes of low-mass hybrid HeCO WD systems, we performed  (3D) hydrodynamical simulations using the moving-mesh code \texttt{Arepo} \citep{2010SpringelArepo, 2011PakmorArepoMHD}. This method allows for accurate modeling of mass transfer dynamics and detonation conditions in interacting WDs. The simulations were coupled to an extended nuclear reaction network to capture the nucleosynthesis driven by helium shell detonations. In general, this interaction of relatively similar mass leads to dynamically strong mass transfer during the merger. However, given the relatively low mass of these WDs, it is not clear whether conditions sufficient to trigger a full thermonuclear explosion will be reached, motivating the exploration of possible merger outcomes in our simulations.

We simulated a binary system consisting of two hybrid HeCO WDs with masses of 0.58 M$_\odot$ and 0.62 M$_\odot$, each containing a CO core surrounded by a substantial helium (He) shell. The initial orbital separation was set to twice the Roche lobe radius—approximately $5.3 \times 10^4$ km, allowing us to resolve the full inspiral and merger phases. 

The initial radial profiles of density and composition for both WDs were constructed following the synthetic models described in
\cite{2022Pakmor, Glanz2025HVWD}, with He shell masses of 0.0365 M$_\odot$ and 0.052 M$_\odot$ for the primary and secondary, respectively. These values are consistent with evolutionary calculations from \cite{Zen+18}, which predict typical helium shell masses resulting from prior binary interaction phases. We adopt these values as representative progenitor structures, and use the simulations to explore the possible merger outcomes.

We follow the inspiral phase and the merger using the same method as we describe in \cite{Glanz2025HVWD}. The nuclear reaction network includes the dominant isotopes formed in helium burning, enabling a comprehensive assessment of nucleosynthetic yields. The system is evolved through mass transfer, detonation, and the subsequent dynamical expansion. The simulation is continued until $\sim100$ s after ignition, by which time nuclear burning has effectively ceased and the ejecta approaches homologous expansion. This duration ensures that both the mass transfer phase and the full development and breakout of the helium detonation are captured self-consistently. Special attention was given to the conditions under which the helium shell detonation would propagate, or fail to propagate, into the CO core. Detailed analysis of the temperature, density, and shock front evolution in the helium shell allowed us to determine the critical conditions for incomplete detonation scenarios.

\section{Results} \label{sec:results}

\begin{figure*}
\includegraphics[width=\textwidth]{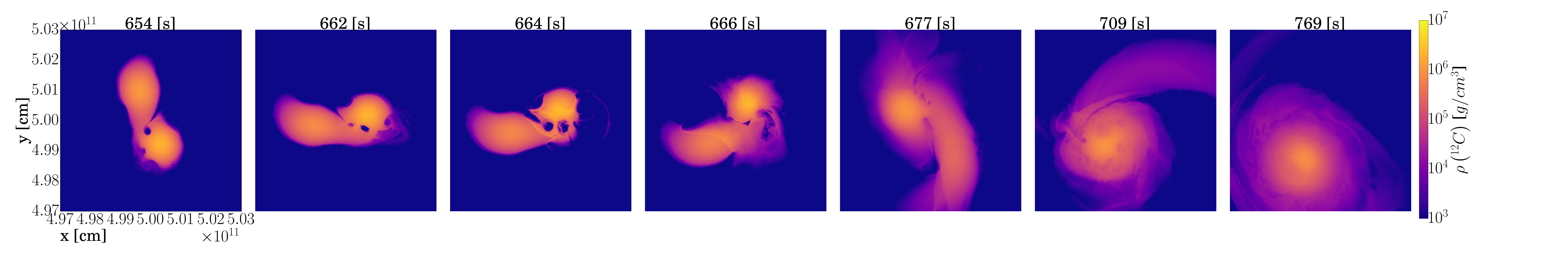}\\
\includegraphics[width=\textwidth]{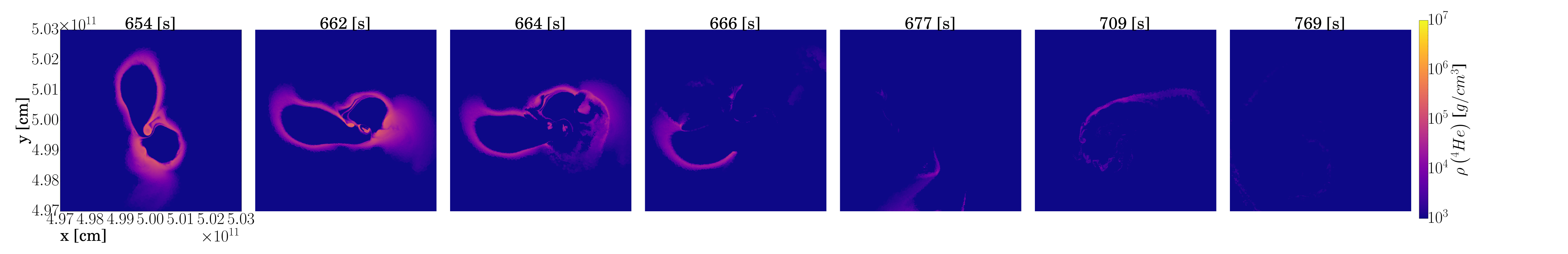}\\
\includegraphics[width=\textwidth]{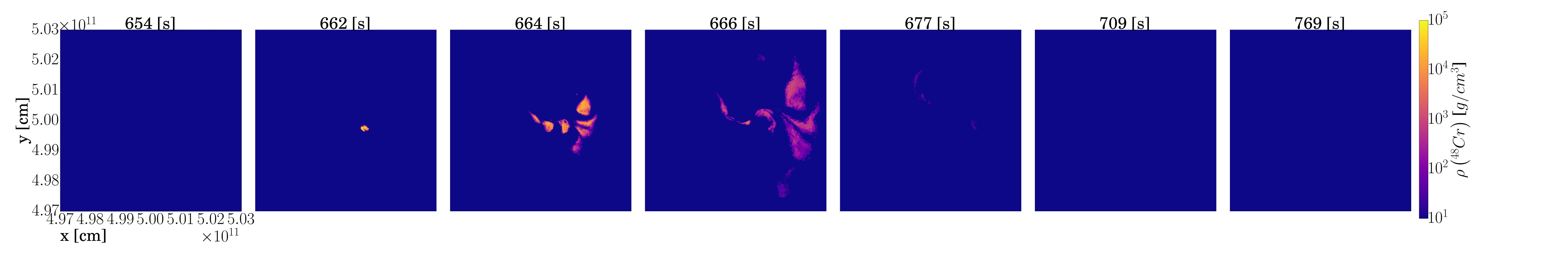}\\
\includegraphics[width=\textwidth]{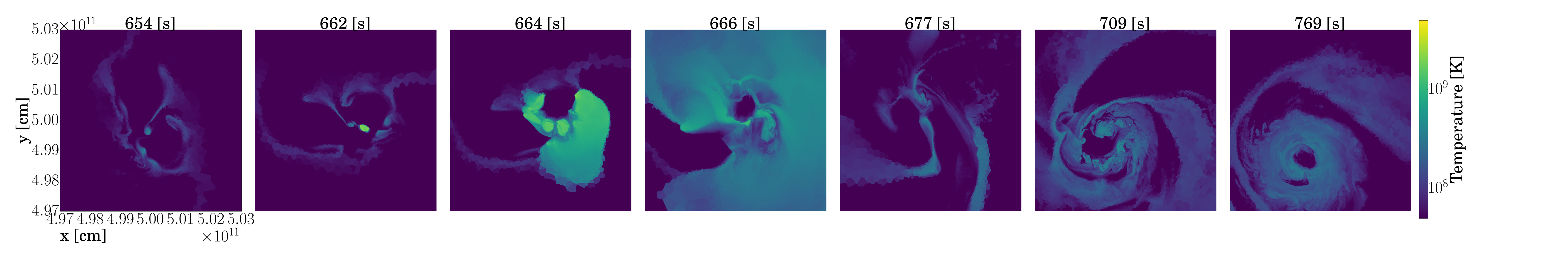}
\caption{The panels show the time evolution from the time of the secondary disruption (left panels), to ignition of the helium (second from left), to the time when the shock converges (middle panels), and finally the ejected remnant in the right panels, still rotating and mixing the bound elements.}
\label{fig:evolution}
\end{figure*}

Our 3D hydrodynamical simulation of the low-mass HeCO WD merger reveals an outcome distinct from that of previously studied higher-mass systems. During the merger, mass transfer from the secondary to the primary triggers a helium shell detonation in the accreted layer on the primary WD. However, this detonation does not propagate into the CO core, leaving the primary WD’s core intact. Figure \ref{fig:evolution} shows the key phases of the system's evolution: the initial tidal disruption of the secondary, ignition of helium in the mass-transfered shell, shock propagation and convergence, and finally the formation of a fast-moving remnant composed of mixed elements. This figure shows how localized hotspots first develop in the helium shell, which is partially mixed with carbon at the surface of the primary WD. These hotspots trigger helium detonation fronts that propagate through the surrounding shell. The lower panels show how the resulting shock travels within the shell and eventually converges; however, the energy released upon convergence is insufficient to ignite a secondary detonation in the underlying CO core. As a result, the CO core remains intact, and the merger produces a partially burned remnant composed of CO and helium-processed material.

\subsection{The supernova}
Approximately 10\% of the total system mass becomes unbound, where the unbound fraction is identified by selecting fluid elements with positive total specific energy (kinetic + internal – gravitational) at the end of the dynamical interaction and summing their mass. Much of this ejected material originates from the disrupted secondary and shock-heated interface during coalescence, and carries intermediate-mass elements and radioactive isotopes such as $^{48}$Cr and $^{52}$Fe. Compared to other WD merger scenarios, this unbound mass fraction is moderate, reflecting the incomplete He shell detonation and partial disruption in these low-mass HeCO pairs. The absence of significant $^{56}$Ni production limits the energy release, consistent with a faint, rapidly evolving transient. 

\begin{figure*}
\centering 
 \includegraphics[width=0.6\linewidth,clip]{ 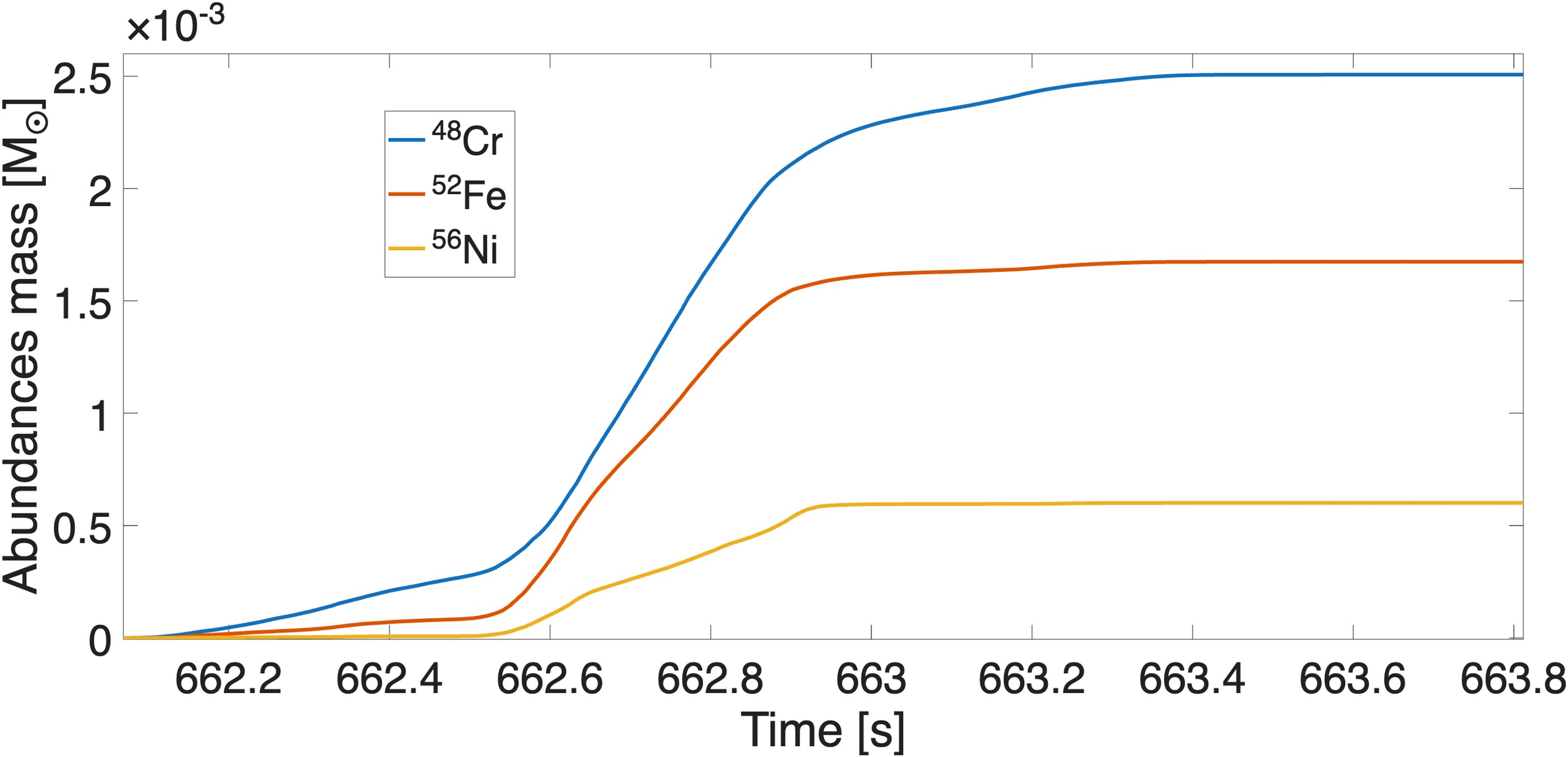}
 \caption{\textbf{Production of radioactive elements throughout the merger.} The incomplete burning of the primary led to the formation of a very low amount of $^{56}$Ni, and more production of lighter elements such as the fast-decaying radioactive elements $^{48}$Cr and $^{52}$Fe that will still produce a luminous transient. Time indicated on the x-axis refers to the time relative to the beginning of the simulation.}
 \label{fig:radioactive}
\end{figure*}

Figure~\ref{fig:radioactive} shows the evolution of newly formed elements following helium ignition. 
We detail the nucleosynthetic yields in Table 1, which provides the masses of iron-group elements and intermediate elements produced during the event. Unlike normal Type Ia supernovae, the early light curves of these fast-evolving events are significantly powered by the decay of short-lived radioactive isotopes such as $^{48}$Cr and $^{52}$Fe, in addition to $^{56}$Ni \citep[e.g.,][]{2010Shen.IaSN}. Based on the total radioactive mass synthesized in our models ($\sim 0.0046 \, \text{M}_{\odot}$), we estimate peak absolute magnitudes on the order of $M \approx -14$ \citep{1982Arnett,2010Shen.IaSN}. Combined with the low ejecta mass, we expect characteristic evolution timescales of a few days. We note, however, that detailed non-local thermodynamic equilibrium (non-LTE) radiative transfer modeling is likely required to provide a more accurate estimation of these observational signatures. These properties indicate that this class of transients could be observable as faint, fast-evolving supernovae, analogous to ``.Ia'' supernovae or Ca-rich gap transients \citep[e.g.,][]{Bildsten_2007, 2010PeretsNature, Waldman2011, Kas+12, Zen+23}, making them particularly suited to high-cadence surveys like the Vera C. Rubin Observatory's LSST.

\begin{table*}
    \centering
    \begin{tabular}{ccc}
        \hline
        Element & Ejected Mass (M$_\odot$) & Bound Mass (M$_\odot$) \\
        \hline
        Total & $0.13$ & $1.07$ \\
        Iron-group (e.g., \(^{48}\text{Cr}\), \(^{52}\text{Fe}\)) & $4.6 \times 10^{-3}$ & $1.6 \times 10^{-4}$ \\
        Intermediate elements & $6 \times 10^{-2}$ & $1.1 \times 10^{-2}$ \\
        \hline
    \end{tabular}
    \caption{Nucleosynthetic Yields from the HeCO WD merger simulation. The iron group stands for Scandium to Nickel, Intermediate: Fluorine to Calcium. Bound material is calculated by finding all mass with a positive energy, including the internal energy, relative to the high-density material at the center of the remnant. The velocity refers to the velocity of the highest central density region of the remnant.}
    \label{tab:yields}
\end{table*}

\subsection{The remnant star}
The surviving primary WD is imparted a recoil velocity of approximately 370kms$^{-1}$ due to asymmetric mass ejection. As shown in the right panel of Figure~\ref{fig:evolution}, the remnant comprises the original primary WD and a portion of the accreted envelope from the disrupted secondary. Its final composition is a mix of helium, intermediate-mass elements, iron-group isotopes, and unburned CO.

The detailed composition of the remnant is presented in Table \ref{tab:final_elements}. Notably, the remnant retains:
\begin{itemize}
\item \num{1.62e-4}~M$_\odot$ of iron-group elements (Scandium through Nickel),
\item \num{1.12e-2}~M$_\odot$ of intermediate-mass elements (Fluorine through Calcium),
\item \num{1.11e-2}~M$_\odot$ of helium
\item \num{1.057}~M$_\odot$ of CO material.
\end{itemize}
The detailed composition of the remnant is presented in Table \ref{tab:final_elements}. The bound remnant is identified by selecting all cells with negative total specific energy, where the total energy includes kinetic, gravitational, and internal components. Elemental masses are computed by summing the mass of each isotope over all bound cells. The remnant velocity is defined as the mass-weighted velocity of the highest-density central region. This procedure follows the methodology described also in \cite{Glanz2025HVWD}.

\begin{table*}

\begin{tabular}{c|c|c}
Element & Remnant Mass $[\text{M}_\odot]$ & Ejecta Mass $[\text{M}_\odot]$\\
\hline
${}^{}$n & \num{4.43e-21} & \num{3.54e-19}\\
${}^{}$H & \num{3.82e-13} & \num{2.66e-08}\\
${}^{4}$He & \num{1.11e-02} & \num{3.62e-02}\\
${}^{11}$B & \num{1.38e-19} & \num{1.60e-19}\\
${}^{12}$C & \num{5.28e-01} & \num{9.15e-03}\\
${}^{13}$C & \num{9.72e-10} & \num{5.61e-11}\\
${}^{13}$N & \num{2.85e-08} & \num{9.66e-09}\\
${}^{14}$N & \num{5.44e-08} & \num{3.33e-08}\\
${}^{15}$N & \num{4.21e-09} & \num{7.49e-10}\\
${}^{15}$O & \num{6.36e-08} & \num{6.98e-08}\\
${}^{16}$O & \num{5.29e-01} & \num{1.71e-02}\\
${}^{17}$O & \num{9.39e-10} & \num{4.38e-10}\\
${}^{18}$F & \num{1.67e-09} & \num{5.11e-10}\\
${}^{19}$Ne & \num{2.97e-09} & \num{2.41e-09}\\
${}^{20}$Ne & \num{6.16e-03} & \num{5.78e-03}\\
${}^{21}$Ne & \num{1.22e-08} & \num{9.36e-09}\\
${}^{22}$Ne & \num{5.70e-10} & \num{3.37e-10}\\
${}^{22}$Na & \num{2.34e-08} & \num{1.79e-08}\\
${}^{23}$Na & \num{2.67e-06} & \num{1.49e-06}\\
${}^{23}$Mg & \num{3.09e-06} & \num{4.13e-06}\\
${}^{24}$Mg & \num{2.50e-03} & \num{6.04e-03}\\
${}^{25}$Mg & \num{7.22e-07} & \num{5.56e-07}\\
${}^{26}$Mg & \num{1.91e-07} & \num{1.45e-07}\\
${}^{25}$Al & \num{3.37e-07} & \num{3.13e-06}\\
${}^{26}$Al & \num{7.51e-06} & \num{7.03e-06}\\
${}^{27}$Al & \num{4.38e-06} & \num{1.46e-05}\\
${}^{28}$Si & \num{1.65e-03} & \num{1.28e-02}\\
${}^{29}$Si & \num{2.10e-06} & \num{9.51e-06}\\

\end{tabular}
\quad
\begin{tabular}{c|c|c}
Element & Remnant Mass $[\text{M}_\odot]$ & Ejecta Mass $[\text{M}_\odot]$\\
\hline
${}^{30}$Si & \num{5.42e-07} & \num{7.19e-06}\\
${}^{29}$P & \num{1.27e-06} & \num{8.03e-05}\\
${}^{30}$P & \num{1.25e-06} & \num{3.32e-06}\\
${}^{31}$P & \num{4.75e-06} & \num{3.70e-05}\\
${}^{31}$S & \num{1.73e-06} & \num{3.41e-05}\\
${}^{32}$S & \num{4.69e-04} & \num{1.27e-02}\\
${}^{33}$S & \num{1.79e-06} & \num{1.09e-05}\\
${}^{33}$Cl & \num{2.86e-08} & \num{2.53e-05}\\
${}^{34}$Cl & \num{2.58e-06} & \num{4.18e-05}\\
${}^{35}$Cl & \num{1.80e-07} & \num{2.05e-05}\\
${}^{36}$Ar & \num{1.21e-04} & \num{6.20e-03}\\
${}^{37}$Ar & \num{5.05e-06} & \num{1.05e-04}\\
${}^{38}$Ar & \num{9.74e-10} & \num{2.24e-07}\\
${}^{39}$Ar & \num{4.37e-13} & \num{2.73e-12}\\
${}^{39}$K & \num{1.69e-07} & \num{2.57e-05}\\
${}^{40}$Ca & \num{1.97e-04} & \num{1.36e-02}\\
${}^{43}$Sc & \num{4.18e-11} & \num{8.00e-09}\\
${}^{44}$Ti & \num{4.40e-05} & \num{2.50e-03}\\
${}^{47}$V & \num{8.47e-09} & \num{2.15e-07}\\
${}^{48}$Cr & \num{6.48e-05} & \num{2.46e-03}\\
${}^{51}$Mn & \num{1.19e-08} & \num{3.14e-07}\\
${}^{52}$Fe & \num{6.22e-05} & \num{1.62e-03}\\
${}^{56}$Fe & \num{1.66e-07} & \num{2.39e-06}\\
${}^{55}$Co & \num{5.48e-09} & \num{1.82e-07}\\
${}^{56}$Ni & \num{3.46e-05} & \num{5.69e-04}\\
${}^{58}$Ni & \num{4.04e-08} & \num{6.23e-07}\\
${}^{59}$Ni & \num{5.35e-08} & \num{7.74e-07}\\
\\
\end{tabular}
\caption{\textbf{Full composition after the end of the simulation ( at 769 s).} The second row is the element mass bound to the high-density region, third row is the unbound mass.}
\label{tab:final_elements}
\end{table*}

\subsection{Long term evolution}\label{subsec:evolution}

\begin{figure}
\centering 
 \includegraphics[width=0.98\linewidth]{ 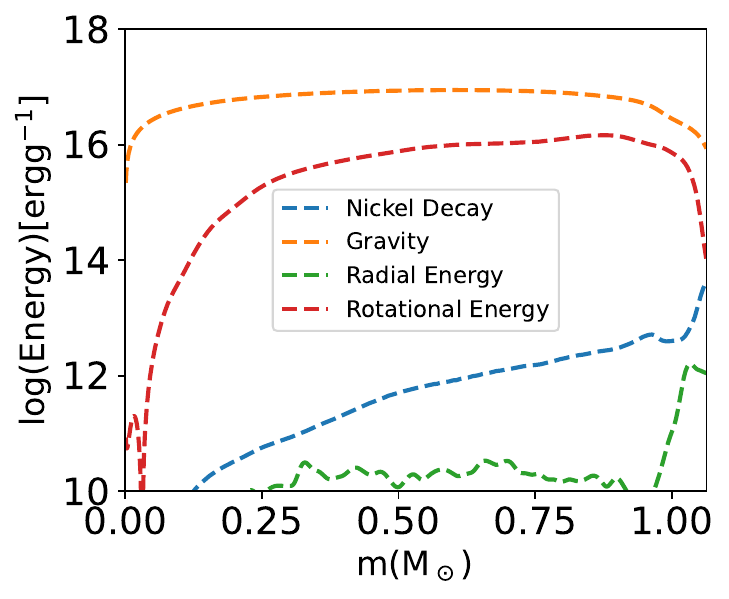}
 \includegraphics[width=0.98\linewidth]{ 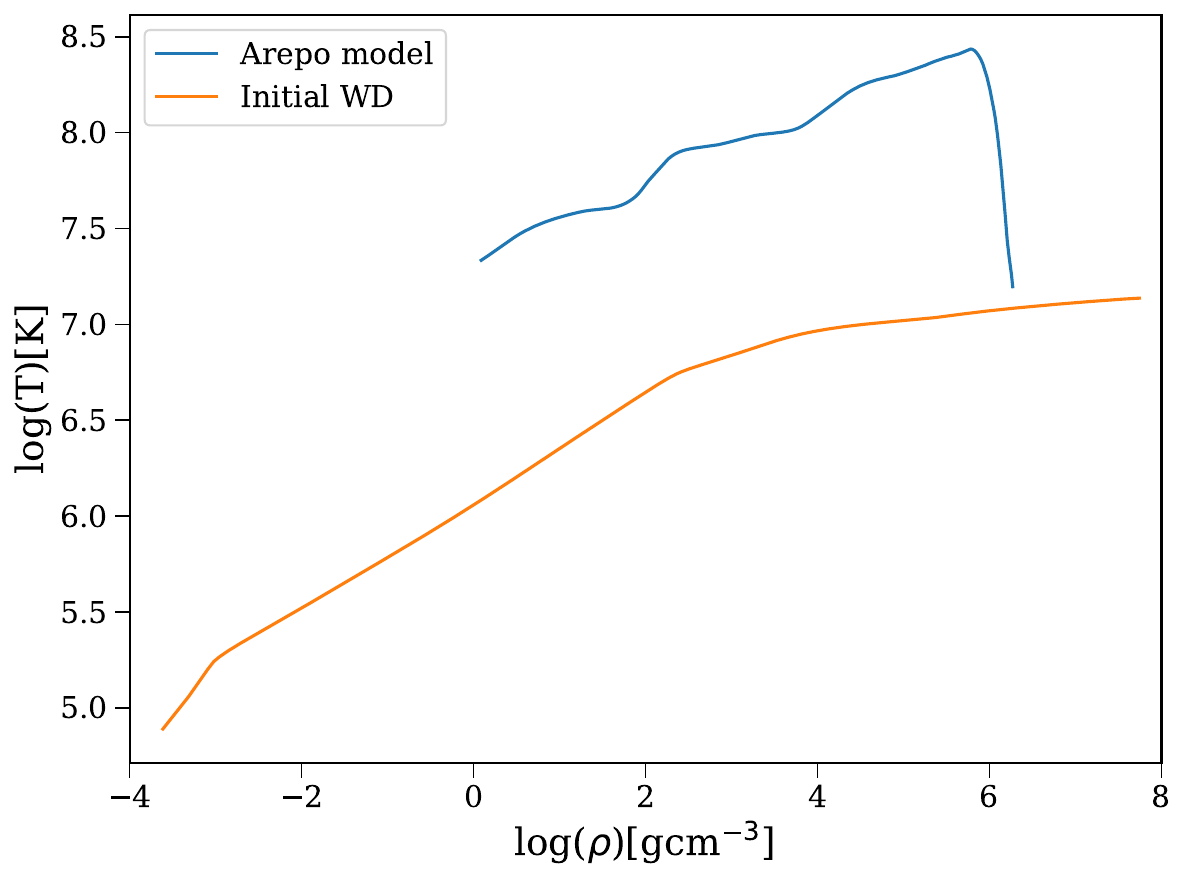}
 \caption{\textbf{Upper panel}: Energy scales of the full model. \textbf{Bottom panel}: Structural profile in $\log\rho-\log{}T$ space for the Arepo model and the initial MESA white dwarf showing the difference in the thermal state.}
 \label{fig:energy}
\end{figure}

The merger of two white dwarfs forms a differentially rotating star which evolves viscously on a timescale of hours \citep{schwab2012}. The post-merger evolution can therefore lead to the heating up of the white dwarf as it settles down to ignite off-center carbon \citep{shen2012}. Here, we do not aim to describe the exact mixing of the remnant and are only concerned with the long-term ($~10^9$ yr) evolution of the merger. As such we follow the remnant in \textsc{Arepo} for $~1000$ seconds until it has dynamically settled and is closer to rigid-body rotation. We follow the further evolution of the remnant using a methodology similar to what was presented in \cite{bhat2024} and \cite{Glanz2025HVWD}. We relax the temperature, density, and angular momentum of the merger product in MESA \citep{2023mesa}. The initial white dwarf is created with the make\_co\_wd test suite and then relaxed to the mass of the remnant. The relaxation of the angular momentum is important here, as the merger rotates quite fast ($~10\%$ of the gravitational energy) but not close to critical (see Figure\ref{fig:energy}). The structure of the WD as shown in the lower panel of Figure\ref{fig:energy} shows that the WD core is surrounded by a hotter, faster rotating envelope. In the previous cases of \citet{bhat2024} and \citet{Glanz2025HVWD}, a portion of the surface was removed prior to the evolution in \textsc{MESA}. This was done primarily to eliminate a small surface layer containing $^{56}$Ni in sufficient quantities that the wind driven by its decay energy caused numerical convergence issues. Since the decay energy exceeded the binding energy of the surface layers, those layers would have been expelled by winds regardless. In this study, we do not make any surface cuts, as the decay energy due to $^{56}$Ni in the atmosphere is insignificant compared to the gravitational energy. We follow the evolution for 1 Gyr, allowing for super-Eddington winds for any mass loss. For the EOS we rely on FreeEOS \citep{Irwin2004} near the surface, and HELM \citep{Timmes2000} and Skye \citep{Jermyn2021} for the cores. Tabulated radiative opacities come primarily from OPAL \citep{Iglesias1993,Iglesias1996}. The low-temperature and high-temperature data are taken from \citet{Ferguson2005} and \citet{Poutanen2017} respectively. The electron conduction opacities are from
\citet{Cassisi2007}. Nuclear reactions are from JINA REACLIB \citep{Cyburt2010}, NACRE \citep{Angulo1999} and additional tabulated weak reaction rates \citet{Fuller1985, Oda1994,Langanke2000}. Screening of reaction rates is included via the prescription of \citet{Chugunov2007}, while thermal neutrino loss rates are from \citet{Itoh1996}. We also adopt the inlist controls from \citet{schwab2021} for the rotational and angular momentum modeling (final inlists will be made available on Zenodo). 

\begin{figure}
\centering 
 \includegraphics[width=0.98\linewidth]{ 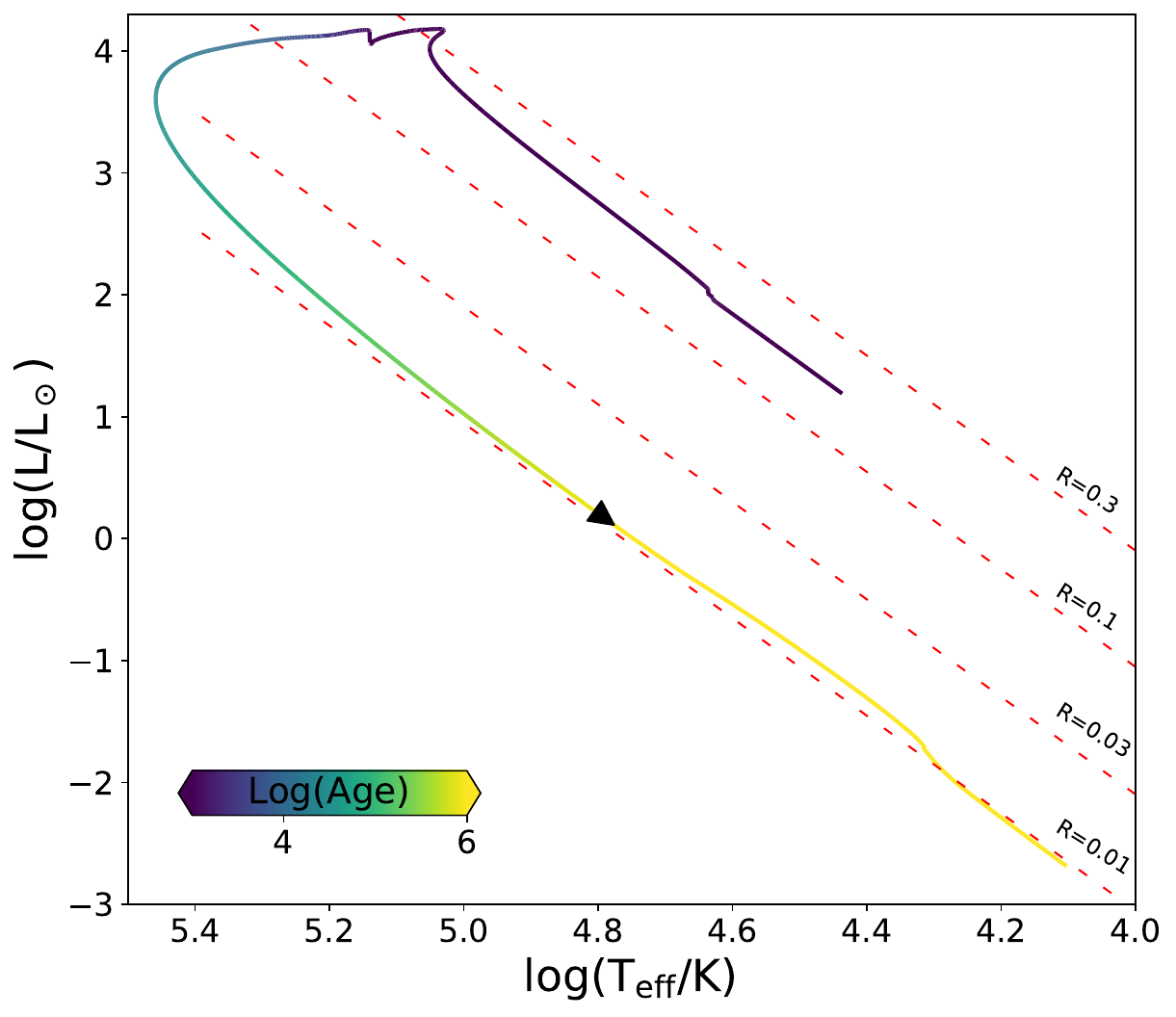}
 \includegraphics[width=0.98\linewidth]{ 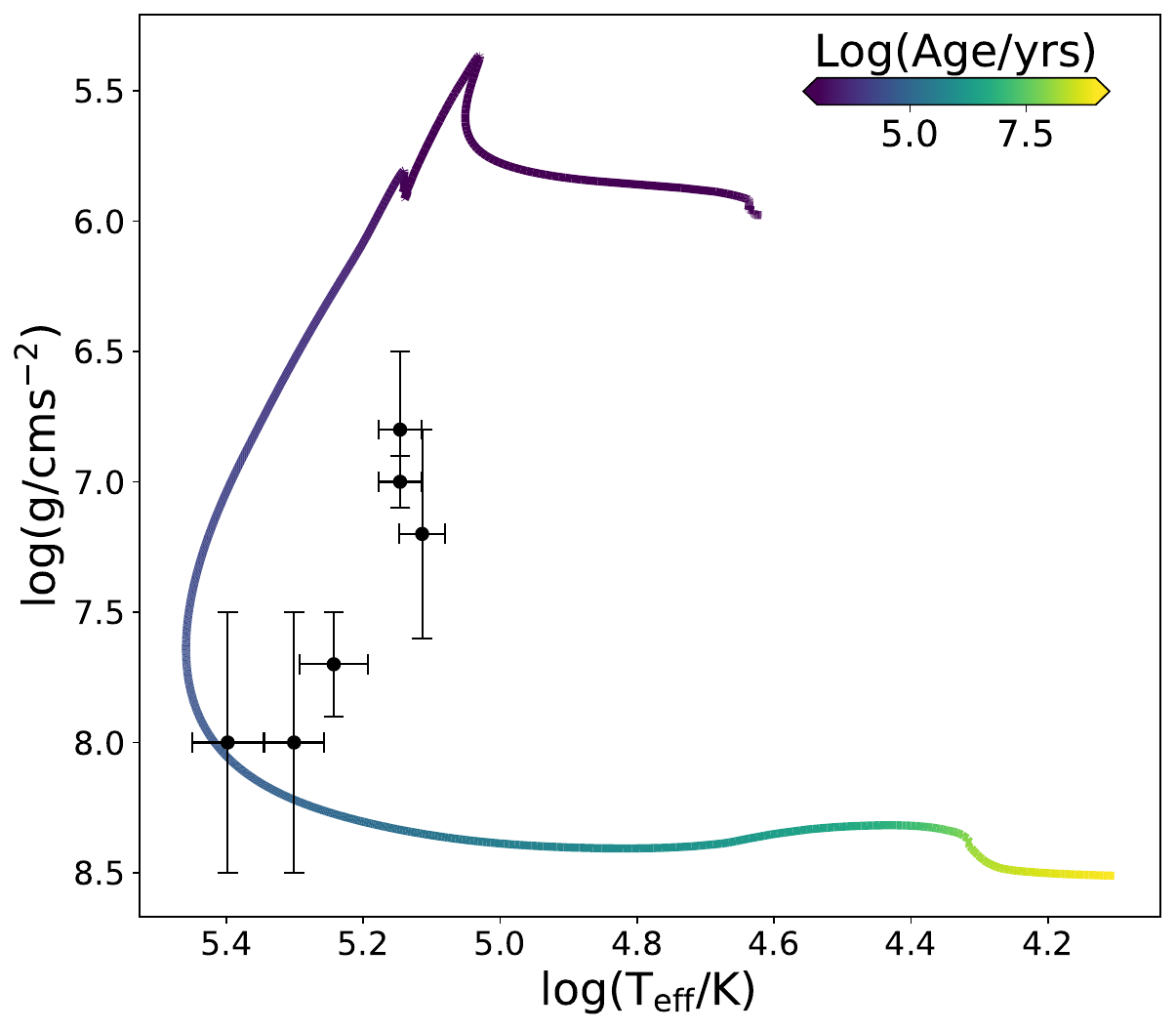}
 \caption{Post-merger evolutionary track in HR and Kiel diagrams. PG1159-type stars are plotted in the Kiel diagram. The black arrow in the HR diagram represents $10^5$ yr. The two hottest stars in the kiel diagram are H1504+65 and RX J0439.8-6809.}
 \label{fig:kiel}
\end{figure}

The evolution of the merger is dominated by radiation pressure for the first $10^5$ Myr. During this time the merger loses around $2\%$ of its mass. This stage also has multiple nuclear processes. The surface is dominated by $^{56}$Ni decay of the order of $10^7$ erg/g/s. The rest of the WD burning comes from triple-alpha process, and alpha capture on carbon (dominant), oxygen, neon, nitrogen, magnesium, and sulphur. There is also a very small amount of carbon burning, of the order of $<10^2$ erg/g/s. The dominant mixing is through thermohaline mixing where we use the prescription from \citet{1980A&A....91..175K}. The mixing smooths out the composition gradient of the heavier elements in the envelope, mixing them into the core. As the $^4$He is used up and the heavier elements sink in nuclear energy becomes negligible around $~1$ Myr, around which time the WD is already on the cooling track. The final WD is a carbon-oxygen WD with a slightly oxygen dominated surface, as expected of massive WDs under 1.06 M$_\odot$ \citep{schwab2021}.


The Hertzsprung-Russel (HR) and Kiel diagrams of the evolutionary track are shown in Figure \ref{fig:kiel}. Our evolutionary tracks suggest an overlap with PG1159 type stars. PG1159 type stars are hot, hydrogen-deficient objects that represent a late stage of stellar evolution before entering the white dwarf cooling track. They are thought to be post-asymptotic giant branch stars. Since the radii of the stars is usually based on mass estimates using evolutionary tracks, we plot PG1159 type stars only in the Kiel diagram to directly compare with spectroscopic results. Plotted stars include the central stars of planetary nebulae, Abell 25 and StDr 138, with values taken from \citet{2024A&A...686A..29W}. We also plot the hot and rapidly-rotating halo star RX J0122.9-7521 \citep{2025A&A...700A..24M} and the prototype of the class, PG1159-035 \citep{2007A&A...462..281J}. We also compare two of the hottest known PG1159 type stars, H1504+65 and its high-velocity twin RX J0439.8-6809 \citep{werner2015}. Both of these stars are carbon and oxygen dominated with neon and magnesium in their spectra, and a deficit of helium and hydrogen. So far no proper evolutionary channel can explain both the abundances and the masses of these two stars. Our model presents an alternative explanation for these elements being present at the surface and explains the observed velocities. However, a more thorough treatment with diffusion must be done before exact surface compositions can be discussed.

\section{Discussion} \label{sec:discussion}

Our findings offer new insights into the diversity of explosive outcomes in double-degenerate WD systems. Unlike mergers involving a CO WD and a massive HeCO WD, where the helium shell detonation often leads to a subsequent carbon detonation in the core, lower-mass HeCO WD pairs exhibit incomplete detonation. This outcome prevents core ignition, resulting in a less energetic, rapidly evolving transient with distinct nucleosynthetic signatures.

The elemental abundances produced in these mergers, particularly the presence of intermediate elements and iron-group isotopes other than \(^{56}\text{Ni}\), suggest that these events would manifest as faint supernovae, potentially resembling observed fast, faint transients. Our results align with observed characteristics of certain unusual supernovae, potentially expanding the known diversity of thermonuclear transients.

These findings also underscore the importance of detailed 3D simulations in capturing the full range of possible outcomes in WD mergers. Previous 2D models were limited in their ability to simulate comparable-mass mergers, where the lack of a stable accretion disk and more complex tidal interactions lead to differing dynamical evolution and detonation outcomes.

\section{Summary} \label{sec:summary}

In this study, we presented 3D hydrodynamical simulations of mergers between two low-mass hybrid HeCO white dwarfs. In addition, we followed the remnant of the merger and evolved it in MESA over a timescale of $1$ Gyr. These systems offer an alternative pathway for thermonuclear explosions that do not reach core detonation, resulting in faint, fast-evolving supernova-like transients. Our key findings include:

\begin{itemize}
    \item Helium shell detonations that fail to propagate into the CO core, leaving the primary WD core intact.
    \item Ejection of intermediate-mass elements and iron-group isotopes (excluding significant \(^{56}\text{Ni}\) production), consistent with faint transient characteristics.
    \item Predicted observable signatures aligned with fast, faint transients, which upcoming surveys like the LSST may detect.
    \item The long-term evolution suggests that hot, fast-rotating, and high-velocity (mainly halo) objects might be products of such mergers. In particular PG1159-type objects dominated by carbon and oxygen in the surface could be explained by this model.
\end{itemize}

These results expand the theoretical landscape of WD mergers and thermonuclear explosions, demonstrating that low-mass HeCO WD systems may contribute to the diversity of observed transients. Our findings highlight the need for further investigation into the initial conditions and mass ratios of double-degenerate HeCO WD.

\section*{Software and third party data repository citations} \label{sec:cite}
We make use of the following codes: 
\begin{itemize}
    \item \texttt{Arepo} \citep{2010SpringelArepo,2020WeinbergArepoPublic,2011PakmorArepoMHD,pak+16}
    \item \texttt{MESA} \citep{2011MESA,Paxton2013, 2015mesa, 2018mesa,2019mesa, 2023mesa}
    \item \texttt{Matplotlib} \citep{Hunter:2007}
    \item \texttt{NumPy} \citep{harris2020arrayNUMPY}
    \item \texttt{SciPy}\citep{2020SciPy-NMeth}
\end{itemize}
\texttt{MESA} inlists will be made available through the \texttt{MESA} community on Zenodo.

\section*{Acknowledgments}
HBP acknowledges support for this project from the European Union's Horizon 2020 research and innovation program under grant agreement No 865932-ERC-SNeX. HG acknowledges support for the project from the Council for Higher Education of Israel.  A.B.
was supported by the Deutsche Forschungsgemeinschaft (DFG)
through grant GE2506/18-1 and by the Kavli Summer Program which took place at MPA in Garching in July 2023, and was supported by the
Kavli Foundation.

\bibliographystyle{aasjournal} 
\bibliography{ms}






\end{document}